\documentclass[trackchanges, floatfix, ApJL, twocolumn, usenames,dvipsnames]{aastex631} 

\usepackage{amssymb}
\usepackage{amsmath}
\usepackage{aas_macros}
\usepackage{color, units}
\usepackage[normalem]{ulem}

\usepackage{xspace}
\usepackage{subfigure}
\usepackage{comment}
\usepackage{xcolor}
\usepackage{bm}
\usepackage{dcolumn}
\usepackage{hyperref}
\usepackage{enumitem} 
\usepackage{array}
\usepackage{nicematrix}
\usepackage{hhline}
\usepackage{graphicx}
\usepackage{multirow}
\usepackage[normalem]{ulem}
\usepackage{soul}
\usepackage[math]{cellspace}
\usepackage{nicematrix} 

\newcommand{\fancytext}[1]{{\relax\ifmmode#1\else $#1$\fi}\xspace}
\newcommand{\mathcmd}[1]{{\sc \relax\ifmmode#1\else $#1$\fi}\xspace}

\newcommand{\xeff}{\mathcmd{\chi_{\text{eff}}}}

\newcommand{\odds}{\mathcmd{\mathcal{O}^{\rm A}_{\rm B}}}
\newcommand{\kdeodds}{\mathcmd{{}^{K}\mathcal{O}^{\rm A}_{\rm B}}}

\newcommand{\SPA}{School of Physics and Astronomy, Monash University, VIC 3800, Australia}
\newcommand{\OzGravMonash}{OzGrav: The ARC Centre of Excellence for Gravitational Wave Discovery, Monash University, VIC 3800, Australia}

\newcommand{\chia}{Chia \textit{et al.}}
\newcommand{\mateu}{Mateu-Lucena \textit{et al.}}
\newcommand{\nitz}{Nitz \textit{et al.}}

\newcommand{\chiap}{Chia${}_{+}$}
\newcommand{\mateup}{ML${}_{+}$}
\newcommand{\nitzp}{Nitz${}_{+}$}

\newcommand{\imrphenomxphm}{\texttt{IMRPhenomXPHM}}

\BeforeBegin{NiceTabular}{\let\begin\BeginEnvironment\let\end\EndEnvironment}
\BeforeBegin{NiceArray}{\let\begin\BeginEnvironment}
\BeforeBegin{NiceMatrix}{\let\begin\BeginEnvironment}

\newcommand{\qMateu}{\mathcmd{{0.52}_{-0.21}^{+0.27}}}
\newcommand{\xeffMateu}{\mathcmd{{0.20}_{-0.06}^{+0.10}}}
\newcommand{\qChia}{\mathcmd{{0.30}_{-0.13}^{+0.22}}}
\newcommand{\xeffChia}{\mathcmd{{0.30}_{-0.11}^{+0.17}}}
\newcommand{\qNitz}{\mathcmd{{0.53}_{-0.23}^{+0.24}}}
\newcommand{\xeffNitz}{\mathcmd{{0.22}_{-0.06}^{+0.10}}}

\newcommand{\MkdeMvC}{\mathcmd{3.45\pm0.01}}
\newcommand{\CkdeMvC}{\mathcmd{0.33\pm0.01}}
\newcommand{\NkdeMvC}{\mathcmd{1.72\pm0.01}}
\newcommand{\priMvC}{\mathcmd{1.68\pm0.01}}
\newcommand{\bfMvC}{\mathcmd{1.0\pm0.2}}
\newcommand{\deepOMvC}{\mathcmd{1.7\pm0.4}}

\newcommand{\MkdeMvN}{\mathcmd{1.46\pm0.04}}
\newcommand{\CkdeMvN}{\mathcmd{0.93\pm0.01}}
\newcommand{\NkdeMvN}{\mathcmd{0.87\pm0.01}}
\newcommand{\priMvN}{\mathcmd{1.08\pm0.01}}
\newcommand{\bfMvN}{\mathcmd{1.3\pm0.2}}
\newcommand{\deepOMvN}{\mathcmd{1.4\pm0.3}}

\newcommand{\MkdeNvC}{\mathcmd{2.38\pm0.1}}
\newcommand{\CkdeNvC}{\mathcmd{0.36\pm0.01}}
\newcommand{\NkdeNvC}{\mathcmd{1.97\pm0.01}}
\newcommand{\priNvC}{\mathcmd{1.56\pm0.01}}
\newcommand{\bfNvC}{\mathcmd{0.8\pm0.2}}
\newcommand{\deepONvC}{\mathcmd{1.2\pm0.3}}

\definecolor{OliveGreen}{rgb}{0,0.6,0}

\newcommand{\newtxt}[1]{#1} 

\begin{document}
\title{Deep follow-up for gravitational-wave inference: a case study with GW151226}
\author{Avi Vajpeyi}
\email{avi.vajpeyi@monash.edu}

\author{Rory Smith}

\author{Eric Thrane}
\affiliation{\SPA}
\affiliation{\OzGravMonash}

\begin{abstract}
New analyses of gravitational wave events raise questions about the nature of some events. 
For example, LIGO--Virgo--KAGRA initially determined GW151226 to be a merger with a mass-ratio $q\approx0.5$ and effective inspiral spin $\chi_{\text{eff}}\approx 0.2$. However, recent works offer an alternative picture: GW151226 is a lower-mass-ratio event $q \approx 0.3$ with slightly higher spin $\chi_{\text{eff}}\approx 0.3$.
This discrepancy has been challenging to resolve as a wide range of differences are employed for each analysis.
This work introduces a ``deep follow-up'' framework to efficiently compute the posterior odds between two different peaks in parameter space.
In doing so, we aim to help resolve disputes about the true nature of gravitational-wave events associated with conflicting astrophysical interpretations. 
Our proposal is not a replacement for standard inference techniques; instead, our method provides a diagnostic tool to understand discrepancies between conflicting results.
We demonstrate this method by studying three {$q$-$\chi_{\text{eff}}$} peaks proposed for GW151226.
We find the $(q\sim0.5, \chi_{\text{eff}}\sim0.2)$ interpretation is only slightly preferred over the $(q\sim0.3, \chi_{\text{eff}}\sim0.3)$ hypothesis with a posterior odds of $\sim1.7\pm0.4$, suggesting that neither of the two peaks can be ruled out. 
We discuss strategies to produce more reliable parameter estimation studies in gravitational-wave astronomy.
\end{abstract}


\section{Introduction}
The binary black hole merger GW151226 was the second gravitational-wave event detected \citep{gw151226}.
The LIGO-Virgo-Kagra (LVK) collaboration \citep{ligo, virgo} measured the event with a signal to noise ratio $\rm{SNR}\sim 13$,  source-frame masses of $(m_1, m_2) = (14.2_{-3.7}^{+8.3} M_\odot, 7.5_{-2.3}^{+2.3} M_\odot$, 90\% credibility)
and with the mass ratio $q \equiv m_2/m_1\approx0.5$~\citep{gw151226}.\footnote{The subscript $1$ refers to the primary, more massive black hole while the $2$ refers to the less massive secondary.}
The event was noteworthy at the time of detection because---unlike GW150914, the first gravitational-wave detection---GW151226 showed signs of black hole spin.
The effective inspiral parameter,
\begin{align}\label{eq:xeff}
    \chi_\text{eff} \equiv \frac{\chi_1\cos\theta_1  + q\chi_2\cos\theta_2}{1+q}
\end{align}
is a measure of the black hole spin projected onto the orbital angular momentum axis.\footnote{$\chi_{1,2}$ are the dimensionless black-hole spins and $\theta_{1,2}$ are the tilt angles between the spin vectors and the orbital angular momentum vector.}
This parameter was found to be inconsistent with zero at high credibility $\xeff=0.18_{-0.08}^{+0.25}$, suggesting that at least one of the black holes in GW151226 was measurably spinning~\citep{gw151226}.
A subsequent analysis by \citet{Mateu-Lucena2021} draws a similar conclusion using a more sophisticated gravitational waveform \citep[\imrphenomxphm,][]{Pratten2021}, which incorporates both higher multipoles and precession, than what was available for the initial analysis.
These results suggest GW151226 is a fairly typical merger; most events in the LIGO--Virgo gravitational-wave transient catalog \citep{gwtc-3} are consistent with $q>0.3$ and approximately $20\%$ exhibit signs of measurable spin \citep{Roulet2021,BuildingBetterModels}. 

However, \citet{Sheng2021} suggest this interpretation of GW151226 is incomplete.
\chia\ reanalyze GW151226 employing a uniform-in-$\xeff$ prior, rather than an isotropic spin prior. 
While the uniform-in-$\xeff$ prior may be astrophysically implausible \citep[Fig 6.]{Galaudage:2021:ApJL}, \chia\ suggest that it is useful for exploring regions of parameter space associated with large values of $\xeff$, which are disfavored by the LVK's isotropic spin prior.
\chia\ find a low-$q$ likelihood peak with source-frame $(m_1, m_2) = (19.1_{-7.1}^{+12.1} M_\odot, 5.8_{-1.7}^{+2.9} M_\odot$) and $q=\qChia$.\footnote{There are other subtle differences, such as different stochastic samplers, different noise estimation techniques, different sampling frequencies, and different methods for likelihood estimation, from previous analyses.}
Due to the degeneracy between $q$ and $\xeff$, the low-$q$ peak is associated with a larger effective inspiral spin $\xeff=\xeffChia$.
This low-$q$ peak differs from the original LVK result \citep{gw151226} and the more recent analysis by \cite{Mateu-Lucena2021}, and \cite{pycbc_ogc_3} which find posterior peaks at $(q,\xeff)=(\qMateu, \xeffMateu)$ and (\qNitz, \xeffNitz) respectively (a high-$q$ peak), although \nitz\ also find weak support for a low-$q$ peak.\footnote{Since this work began, \nitz\ published updated posteriors for GW151226 in \cite{pycbc_ogc_4}, with a {$q$-$\xeff$} posterior peak at $(0.52^{-0.2}_{+0.26}, 0.21^{-0.06}_{+0.1})$.} 

The low-$q$ interpretation makes GW151226 a more interesting event.
With an asymmetric mass ratio $q\sim3$, GW151226 is less like typical events in GWTC-3~\citep{gwtc-3} and more like the mysterious GW190814~\citep{GW190814}---an event which is difficult to explain in the context of standard binary evolution.
To simultaneously achieve such a large value of $\xeff$ and a small value of $q$, the primary black hole must have non-zero spin \citep{Qin2022}.
However, this is somewhat challenging to explain in the usual framework of tidal interactions, which serve to spin up only the second-born (typically less massive) black hole \citep{Mandel2020}.\footnote{In some events, large primary spin can arise due to the phenomenon of ``mass ratio reversal'' \citep{Olejak2021}, though, this seems unlikely to have occurred in a system with such a small mass ratio.}
Finally, \chia\ suggest that GW151226 exhibits clear signs of orbital precession, making this the first system to exhibit this effect~\citep{Green:2021:PhRvD, Abbott:2021:ApJL}.

While the \chia\ state that the low-$q$ peak likelihood is preferred over the high-$q$ peak likelihood by a factor of $e^2$, \mateu\ and \nitz\ do not find such support for the low-$q$ peak (although they use different spin priors). 
This creates a predicament: the community has two conflicting interpretations of the GW151226 event.
This work demonstrates a method to resolve the tension between two dissimilar sets of parameter estimation results.
This method allows us to compute odds comparing two points in parameter space. 
These odds can be used to ascertain which of two or more inference results are more probable, helpful in understanding discrepancies like those described above.
We apply this technique to GW151226 to settle the debate on the relative importance of the low-$q$ and high-$q$ likelihood peaks. 
We compute the marginal likelihood of two points in the {$q$-$\xeff$} distribution, one from \chia's low-$q$ likelihood peak and another from the \mateu's high-$q$ peak. 
By restricting the analysis to the two-dimensional {$q$-$\xeff$} space, we reduce the problem's dimensionality, permitting a ``deep follow-up'' investigation on the {$q$-$\xeff$} modes of GW151226 without the need for large computational resources.
While we demonstrate this method on GW151226, the same principles can be applied to resolve other discrepancies in the gravitational-wave astronomy literature.
The remainder of this letter is organized as follows. 
In Section~\ref{sec:method}, we describe the ``deep follow-up'' formalism used to compute our posterior odds. 
Then, in Section~\ref{sec:results}, we show results obtained from a deep follow-up of GW151226. 
We provide closing comments in Section~\ref{sec:discussion}.

\section{Methodology}\label{sec:method}

Different analyses may find different peaks in the marginal likelihood distribution. 
This section provides a mathematical method to determine which peak is more probable than another.
We label the two disparate likelihood peaks $A$ and $B$ and ask the question: what are the posterior odds of hypothesis $A$ over hypothesis $B$?
If this odds, denoted $\mathcal{O}^A_B$, is much larger (or much smaller) than unity ($\ln\mathcal{O}\gg 8$ or $\ln\mathcal{O}\ll -8$), then one hypothesis can be said to be strongly preferred over the other.
When this happens, we may conclude that one hypothesis provides a better description of the posterior distribution.\footnote{More precisely, one hypothesis is preferred over the other given whatever assumptions we have made constructing our likelihood function.}
On the other hand, if $\mathcal{O}\approx 1$, this implies that both hypotheses are approximately equally consistent with the data.
When this happens, we may conclude that one or more analysis is incomplete insofar as they did not produce a posterior consistent with both $A$ and $B$.

In the case of GW151226, we choose $A$ and $B$ from \mateu\ and \chia\ peaks in the $(q, \xeff)$ plane because different peaks in this plane can lead to unique physical interpretations of the event. 
As we are only interested in this plane, the marginal likelihood distribution $\mathcal{Z}(d|q,\xeff)$ allows us to focus on just the $2$ parameters at a time; the other parameters (e.g., total mass, the effective precession parameter, etc.) are integrated out.
By fixing the parameter values associated with points $A$ and $B$ we reduce the dimensionality of our inference calculation, which improves convergence; for our GW151226 analysis, the dimensionality of the marginalized likelihood is reduced by two.
Moreover, we do not have to worry about our stochastic sampler failing to find peaks $A$ and $B$---we force the sampler to study just these two points in parameter space.
By setting up the problem this way, we give up trying to map the full $(q, \xeff)$ subspace, focusing our computational power so that we obtain reliable estimates of the marginal likelihood at just the two points $A$ and $B$.
If the inference calculations by \cite{Mateu-Lucena2021} and \cite{Sheng2021} are analogous to all-sky searches, this analysis is analogous to a deep follow-up study.

The posterior odds are 
\begin{align}\label{eq:odds}
    \odds = \frac{Z(d|A)}{Z(d|B)}\, \frac{\pi(A)}{\pi(B)},
\end{align}
where $Z(d|...)$ refers to the marginal likelihood for hypothesis $A$ or $B$ and $\pi(...)$ refers to the prior odds for hypothesis $A$ or $B$. 
The marginal likelihood for point $A$ or $B$ is given by
\begin{align}\label{eq:Z}
    \mathcal{Z}(d|A,B) = \int \mathcal{L}(d|\theta)\, \pi(\theta|A,B),
\end{align}
where $\mathcal{L}(d|\theta)$ is the usual Whittle likelihood used in gravitational-wave astronomy (see, e.g., \cite{intro_to_gw_bayes}) and $\pi(\theta|A,B)$ is the prior on $\theta$ conditioned upon hypothesis $A,B$. 
We first discuss how to calculate the marginal likelihood and then describe how to calculate the prior odds.

\textit{The marginal likelihood.}
Our first step is to define the priors conditioned on hypotheses $A,B$.
In this study, $\pi(\theta|A,B)$ is the isotropic spin prior from \cite{Mateu-Lucena2021}, conditioned on $(q, \chi_\text{eff})_{A,B}$, the location of either \mateu's {$q$-$\xeff$} peak $A$ or \chia's {$q$-$\xeff$} peak $B$.
The isotropic spin prior is uniform in the dimensionless spin parameters $\chi_1, \chi_2$, the azimuthal angle of the spin vectors $\phi_1, \phi_2$, and the cosine of the spin-vector zenith angle $z_1, z_2$.
The prior distribution of black hole masses is uniform in $(m_1, m_2)$ over the interval $(3 M_\odot, 54 M_\odot)$ subject to the constraint that $m_2 < m_1$.
We employ a uniform-in-comoving volume prior for luminosity distance and assume standard priors for the other extrinsic parameters.

The next step is to condition the isotropic spin prior on $(q, \xeff){A,B}$.
Using Eq.~\ref{eq:xeff}, we can express our prior on $\xeff$ as delta function:
\begin{align}
    \pi(\xeff&| \chi_1, z_1, z_2, \chi_2, q_{A,B}) 
    = \delta\bigg( \xeff - \frac{\chi_1 z_1 + q_{A,B} \chi_2 z_2}{1+q_{A,B}} \bigg)\ .
\end{align}
Here, $q_{A,B}$ refer to the mass ratio associated with either hypothesis $A$ or hypothesis $B$.
By changing variables, we obtain a conditional prior on $\chi_2$ for hypotheses $A$ and $B$:
\begin{align}\label{eq:delta_chi2}
    \pi(\chi_2 &| 
        \chi_1, z_1, z_2)_{A,B} 
        =  \delta\left(
        \chi_2 -\frac{\chi_\text{eff}^{A,B} (1+q_{A,B}) - \chi_1 z_1}
        {q_{A,B} z_2}
        \right).
\end{align}
Combining our isotropic spin prior with Eq.~\ref{eq:delta_chi2}, we obtain the marginal prior distribution for $\chi_1$ conditioned on $A,B$\footnote{Probability densities in equations \ref{eq:chi1},\ref{eq:z1}, \ref{eq:z2} and \ref{eq:prior_calc} are normalized numerically.}
\begin{align}\label{eq:chi1}
    \pi(\chi_1)_{A,B}  
    =& 
    \int\, dz_1\,  dz_2\, d\chi_2 \, 
    \pi(\chi_1, z_1, z_2, \chi_2)_{A,B} \nonumber
    \\
     =& 
    \int\, dz_1\,  dz_2\, d\chi_2 \, 
    \pi(\chi_1, z_1, z_2)\, \pi(\chi_2 |\chi_1, z_1, z_2)_{A,B}\nonumber
    \\
    \propto & \frac{1}{n} \sum_{k}^n 
    \mathcal{I}
    \left(0 < \frac{\chi_\text{eff}^{A,B} (1+q_{A,B}) - \chi_1 z_1^k}
    {q_{A,B} z_2^k} < 1\right) 
\end{align}
To obtain the final expression, we wrote the integrals over $z_1, z_2$ as a sum over prior samples drawn from the $\pi(z_1)$ and $\pi(z_2)$ distributions.
The symbol $\mathcal{I}$ denotes an indicator function, which returns zero when its argument is false and one when its argument is true. 
Equation~\ref{eq:chi1} can be used to generate random samples of $\chi_1$ with inverse transform sampling conditioned on either hypothesis $A$ or hypothesis $B$.
We denote these samples $\{\chi_1^j\}_{A,B}$.

\begin{figure}
  \includegraphics[width=\linewidth]{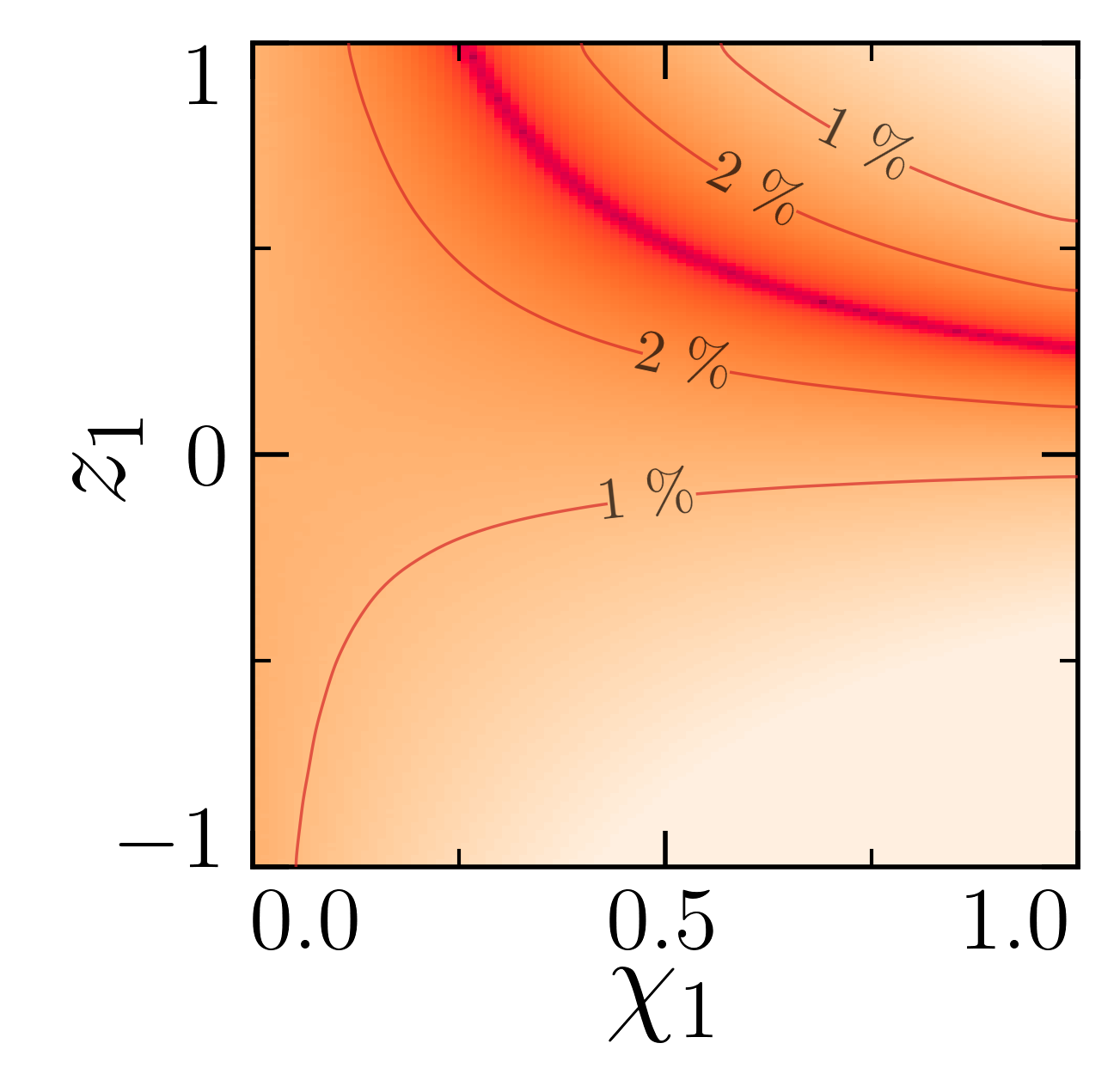}
  \caption{
  Heatmap of $\pi(z_1, \chi_1|q,\xeff)$. Contours are drawn at the $1\%$ and $2\%$ percentiles.
  Here, $(q,\xeff)=(0.68, 0.15)$.
  }
  \label{fig:joint_prior}
\end{figure}

The next step is to add values of $z_1$ to the $\chi_1$ samples.
The marginal prior distribution for $z_1$ conditioned on $\chi_1^j$ and hypothesis $A$ or $B$ is
\begin{align}\label{eq:z1}
    \pi(z_1 &| \chi_1^j)_{A,B} \nonumber \\
    \propto & \frac{1}{n} \sum_{k}^n 
    \mathcal{I}
    \left(0 < \frac{\chi_\text{eff}^{A,B} (1+q_{A,B}) - \chi_1^j z_1}
    {q z_2^k} < 1\right) .
\end{align}
This time, the sum over $k$ is a sum over prior samples for $z_2$.
We use this distribution to generate random samples $z_1^j$ given $\chi_1^j$ (see Fig.~\ref{fig:joint_prior}).\footnote{We compute the distribution in Eq.~\ref{eq:z1} on a two-dimensional grid in $(z_1, \chi_1)$. We perform inverse-transform sampling using columns from this grid.}
Using the same logic, the prior distribution for $z_2$ conditioned on $\chi_1^j, z_1^j$ and hypothesis $A$ or $B$ is
\begin{align}\label{eq:z2}
    \pi(z_2 &|  \chi_1^j, z_1^j)_{A,B} 
    \propto& 
    \mathcal{I}
    \left(0 < \frac{\chi_\text{eff}^{A,B} (1+q_{A,B}) - \chi_1^j z_1^j}
    {q z_2} < 1\right)  .
\end{align}
This distribution allows us to generate a random sample $z_2^j$.
For each sample, we now have $\xeff, q, \chi_1^j, z_1^j, z_2^j$, so the value for $\chi_2$ is completely determined by Eq.~\ref{eq:xeff}.

Using these equations (\ref{eq:chi1}-\ref{eq:z2}), we define priors for our stochastic sampler that are conditioned on hypotheses $A$ and $B$.\footnote{We check the calculations by generating corner plots of our prior samples conditioned on $A,B$ and comparing them to corner plots created with the isotropic spin prior, but keeping only samples in the vicinity of $(q_{A,B},\chi_\text{eff}^{A,B})$.}
We perform two parameter estimation runs, one for hypothesis $A$ and one for $B$.
The evidence from each run yields the marginal likelihood defined in Eq.~\ref{eq:Z}.
The ratio of these two evidence values gives us the Bayes factor
\begin{align}
    \text{BF} = \frac{\mathcal{Z}(d|A)}{\mathcal{Z}(d|B)} ,
\end{align}
which represents the relative probability for hypothesis $A$ over $B$ given even prior odds.
However, the isotropic spin prior described above does not assign even prior odds; small values of $\xeff$ are more likely than larger ones.
Thus, to quantify the posterior odds, we now move on to the prior odds: $\pi_{A,B} = \pi(q_{A,B},\chi_\text{eff}^{A,B})$.
Note that we are not addressing whether the isotropic spin prior is a good (astrophysically-motivated) prior. 
Our goal here is to compute the posterior odds given the isotropic spin prior.

\textit{The prior odds.}
The marginal prior probability density for $(q, \xeff)$ can be calculated like so:
\begin{align}\label{eq:prior_calc}
    \pi(q, \xeff) \propto \frac{2}{n}\sum_k^n 
    \left|\frac{1+q\xeff}{\chi_1^k}\right|
    \left(\frac{\chi_1^k +q\chi_2^k z_2^k}{1+q}\right) .
\end{align}
Here, the sum over $k$ is a sum over isotropic spin prior samples for $(\chi_1, \chi_2, z_2)$.
What follows is a derivation of Eq.~\ref{eq:prior_calc}.
To begin, we note that
\begin{align}
    \pi(q,\xeff) &= \pi(\xeff|q)\, \pi(q) .
\end{align}
We rewrite the first term in terms of marginalization integrals to obtain
\begin{align}\label{eq:marg}
    \pi&(q,\xeff) = \nonumber \\ 
    &\int d\chi_1\,  d\chi_2\, dz_2 \, 
    \pi(\xeff|q, \chi_1, \chi_2, z_2) \,
    \pi(q, \chi_1, \chi_2, z_2) .
\end{align}
We have a closed-form expression for the isotropic spin prior $\pi(q, \chi_1, \chi_2, z_2)$.
The remaining term in Eq.~\ref{eq:marg} is related to the fiducial distribution for $z_1$.
Since $z_1$ is uniformly distributed, $\xeff$ is uniformly distributed with limits depending on $(\chi_1, q, \chi_2, z_2)$:
\begin{align}\label{eq:xeff_cond}
    \pi(\xeff&|q, \chi_1, \chi_2, z_2 ) \nonumber\\
    &= \bigg|\frac{\partial z_1}{\partial \xeff} \bigg|\, U\bigg[\frac{-\chi_1 + q \chi_2 z_2}{1+q}, \frac{ \chi_1 + q \chi_2 z_2}{1+q}\bigg]\, \nonumber\\
    & = \bigg|\frac{1+q \xeff}{\chi_1} \bigg|\, U\bigg[\frac{-\chi_1 + q \chi_2 z_2}{1+q}, \frac{ \chi_1 + q \chi_2 z_2}{1+q}\bigg].
\end{align}
Plugging Eq.~\ref{eq:xeff_cond} into Eq.~\ref{eq:marg} and replacing the marginalization integrals with a sum over prior samples, we obtain Eq.~\ref{eq:prior_calc}.

\begin{figure}
  \includegraphics[width=\linewidth]{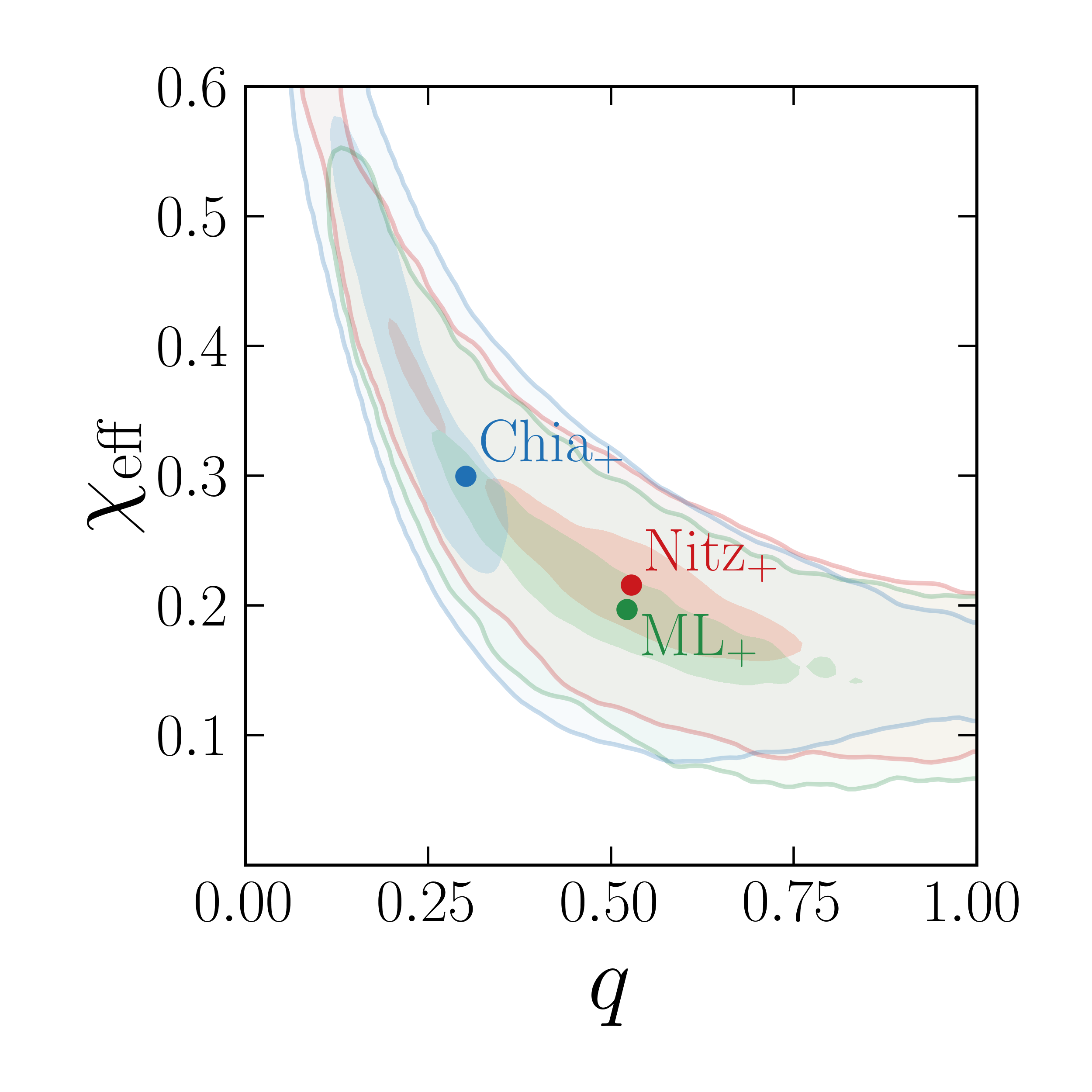}
  \caption{
  GW151226 {$q$-$\xeff$}  posteriors from \citet{Mateu-Lucena2021}~\mateup, \citet{Sheng2021}~\chiap, \citet{pycbc_ogc_3}~\nitzp\ analyses. Contours are drawn at the {1,3-$\sigma$} regions. The points mark the median {$q$-$\xeff$} values of each analysis.
  }
  \label{fig:qxeff_samples}
\end{figure}

\textit{Sampling.}
For hypothesis $A$ (the high-$q$ peak), we set $(q,\xeff)=(0.52, 0.2)$.
For hypothesis $B$ (the low-$q$ peak), we set $(q,\xeff)=(0.3,0.3)$.
These values are the median {$q$-$\xeff$} values from the \mateu\ and \chia\ posteriors and are plotted in  Figure~\ref{fig:qxeff_samples}. 
We compute the marginalised likelihood for each hypothesis using \texttt{Parallel Bilby}~\citep{bilby, pbilby_paper}, employing \texttt{Dynesty}~\citep{dynesty_paper} as our nested sampler~\citep{skilling2004, skilling2006}. 
We use $2,000$ live points, a data sampling rate of $\unit[4096]{Hz}$ and the \newtxt{\imrphenomxphm} waveform.
The configuration files for our analyses are located on this project's GitHub repository~\citep{deep_followup_github}.

\textit{Validation.}
To validate this method, we compute the {$q$-$\xeff$} deep follow-up posterior odds \odds\ for two additional cases: a GW151226-like signal injected in Gaussian noise and GW150914, displayed in Table~\ref{tab:validation}. 
We apply a kernel density estimate (KDE) of the two signals' posterior samples in the $(q, \xeff)$ plane. 
The \odds\ fall within the uncertainty of the \kdeodds, demonstrating that the deep follow-up method yields consistent posterior odds with KDE posterior odds for the signals. Details of the injection and analyses can be found in the same repository~\citep{deep_followup_github}.

\begin{table}
  \centering
  \caption{Comparisons of the deep follow-up posterior odds \odds to kernel density estimate posterior odds \kdeodds for a GW151226-like simulated signal and GW150914.}
  \label{tab:validation}
  \renewcommand{\arraystretch}{1.4}
  \begin{NiceTabular}{ccc}
      \CodeBefore
    \rowcolor{gray!10}{2-3}
    \Body
                  & Simulated     & GW150914      \\ \hline
  $A\, (q,\xeff)$ & $0.36, 0.17$  & $0.7, 0.02$   \\    
  $B\, (q,\xeff)$ & $0.39, 0.25$  & $0.78, -0.04$ \\
  \odds           & $0.47\pm0.11$ & $0.24\pm0.1$  \\
  \kdeodds        & $0.47\pm0.02$ & $0.24\pm0.01$ 
  \end{NiceTabular}
\end{table}

\section{Results}\label{sec:results}

Our deep follow-up findings for GW151226 are summarized in Table~\ref{tab:gw151226-result}.
Assigning the high-$q$, \mateu\ {$q$-$\xeff$} peak, to hypothesis $A$ and the low-$q$, \chia\ {$q$-$\xeff$} peak, to hypothesis $B$, we obtain a Bayes factor of $\bfMvC$, implying that---if we assign even prior odds (the uniform-in-$\xeff$ prior odds) to hypotheses $A$ and $B$---both peaks are equally preferred.
However, the isotropic spin prior odds are $\priMvC$ in favor of the hypothesis $A$.
Thus, we obtain a posterior odds $\odds=\deepOMvC$, implying that high-$q$ peak is only slightly preferred over low-$q$ peak. 
Comparing the \nitz\ and \chia\ {$q$-$\xeff$} peaks yields similar results: a slight preference of the high-$q$ peak over the low-$q$ peak after incorporating the isotropic spin prior odds.
To compare the deep follow-up odds with the \mateu's {$q$-$\xeff$} posterior, we apply a kernel density estimate (KDE) of their posterior samples in the $(q, \xeff)$ plane.
The KDE yields a posterior odds of $\kdeodds=\MkdeMvC$, a value twice the deep follow-up \odds. This suggests that---for whatever reason---the \mateu\ analysis underestimates the posterior support for the \chia\ {$q$-$\xeff$} peak (by a factor of $\approx 2$).
At the same time, using a KDE of the \chia\ samples, we find the \kdeodds between the two points is $\kdeodds=\CkdeMvC$, a value one-fifth the \odds. This suggests that the deep follow-up analysis finds less support for the \chia\ {$q$-$\xeff$} peak than reported by \chia\ (by a factor of $\approx 1/5$). 

\begin{table}
  \centering
  \caption{
  The median {$q$-$\xeff$} values from \citet{Mateu-Lucena2021} \mateup, \citet{Sheng2021} \chiap, and \citet{pycbc_ogc_3} \nitzp\ analyses, along with the prior odds $\pi(A)/\pi(B)$, Bayes-Factors BF, ``deep follow-up'' posterior odds \odds, and kernel density estimate (KDE) posterior odds \kdeodds. The three rows of KDE posteriors odds are computed using the \mateup, \chiap, and \nitzp\ {$q$-$\xeff$} samples. 
  }
  \label{tab:gw151226-result}
  \renewcommand{\arraystretch}{1.1}
  \begin{NiceTabular}{cccc}
      \CodeBefore
    \rowcolor{gray!10}{2-3}
    \rowcolor{gray!10}{7-9}
    \Body
                         & \Block{}{\mateup vs \\ \chiap}          & \Block{}{\mateup vs \\ \nitzp}            & \Block{}{\nitzp vs \\ \chiap}       \\ \hline            
  $A\, (q,\xeff)$        & $0.52, 0.20$         & $0.52, 0.20$        &  $0.53, 0.22$           \\    
  $B\, (q,\xeff)$        & $0.30, 0.30$         & $0.53, 0.22$        &  $0.30, 0.30$           \\ 
  $\pi(A)/\pi(B)$        & \priMvC              & \priMvN             &  \priNvC                \\
  BF                     & \bfMvC               & \bfMvN              &  \bfNvC                 \\ 
  \odds                  & \deepOMvC            & \deepOMvN           &  \deepONvC              \\ 
  \mateup \ \kdeodds     & \MkdeMvC             & \MkdeMvN            &  \MkdeNvC               \\
  \chiap \ \kdeodds      & \CkdeMvC             & \CkdeMvN            &  \CkdeNvC               \\
  \nitzp \ \kdeodds      & \NkdeMvC             & \NkdeMvN            &  \NkdeNvC               \\
  \end{NiceTabular}
\end{table}

\section{Discussion}\label{sec:discussion}
Increasingly, analyses of gravitational-wave events have contradicting results (e.g., GW151226, GW190425, GW190521 \citealt[]{gwtc-3, ias_o3a_catalog, pycbc_ogc_3}).
Faulty inferences of events can provide misleading interpretations of the population.
In some cases, these analyses differ in priors, software, sampler settings, data, waveforms, and noise models, making the determination of the ``correct'' analysis challenging. 
The method to compute posterior odds demonstrated in this work can help us understand contradictory results, allowing the community to determine the relative preference of one point over another in parameter space.

This work investigates the contradicting interpretations for GW151226 between \cite{Sheng2021} and other analyses. 
\cite{Sheng2021} suggest that the low-$q$ mode arises from their usage of the uniform-in-$\xeff$ spin prior. 
Although \citet{Galaudage:2021:ApJL} argue that the uniform-in-$\xeff$ is astrophysically unrealistic, \citet{Sheng2021} state that this prior does not suppress the large $\xeff$\ regions of parameter space.

In addition to the difference in priors used by \cite{Sheng2021} and other analyses, \citet{Sheng2021} use a different method to estimate the noise power spectral density (PSD), the likelihood, and even conduct different data cleaning. 
Our study's output can help study the impact of the different techniques on this event.
For example, one can study the impact of an alternative PSD by re-weighting the posterior samples for points $A$ and $B$ (publicly accessible at~\citet{deep_followup_github}) with a different PSD. The steps for this are described in Appendix~\ref{sec:appendix}.

\begin{figure*}
\begin{center}
  \includegraphics[width=.7\linewidth]{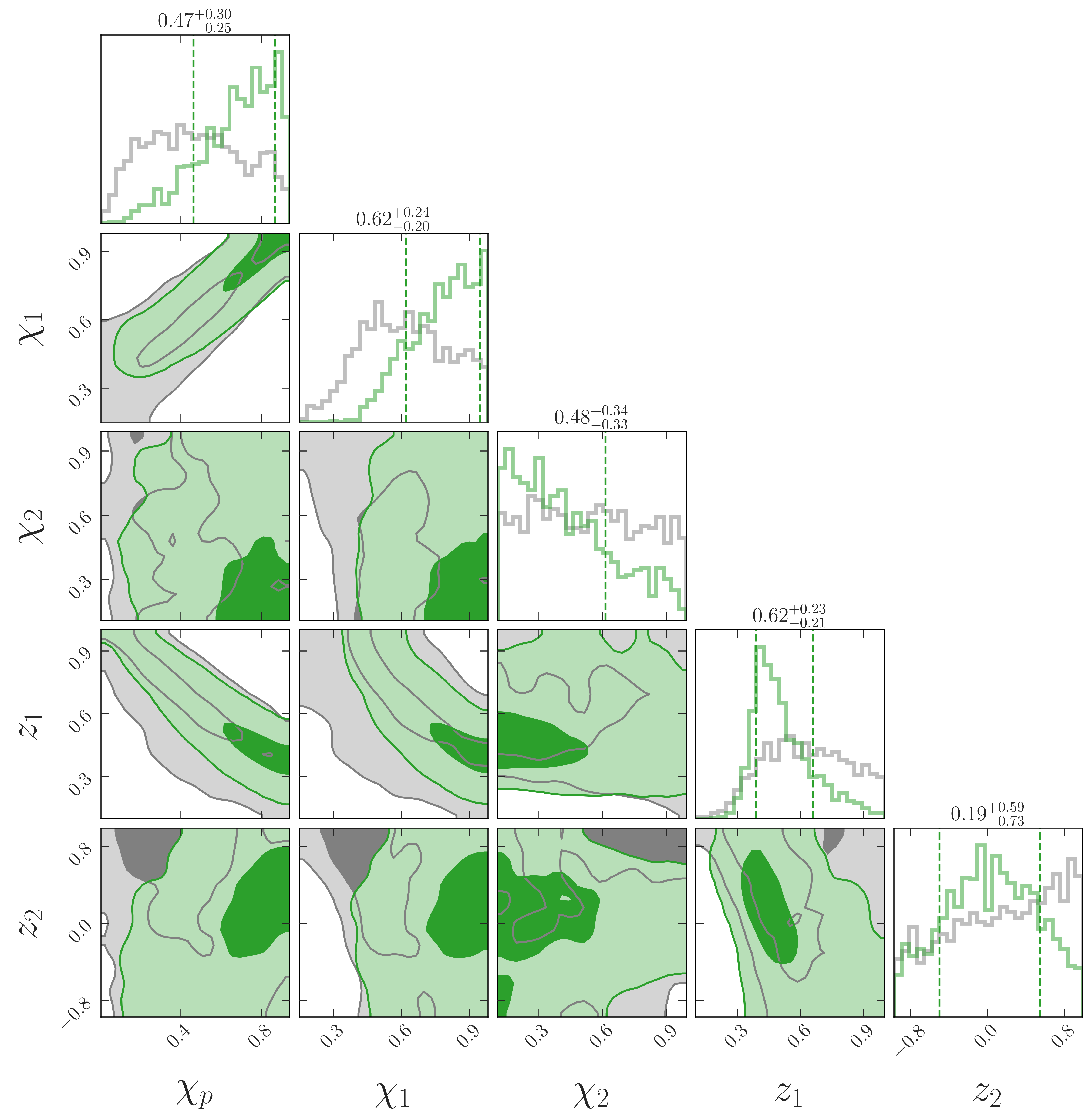}
  \caption{
    Posterior (green) and prior (gray) samples from the low-$q$ (\chia) deep follow-up analysis. The contours are drawn along the {1,3-$\sigma$} regions. The vertical dashed lines in the 1D marginal distributions are plotted at the {1-$\sigma$} credible regions for the posterior. The median and {1-$\sigma$} credible values for the posterior's marginal distributions are included above the 1D plots. 
  }
  \label{fig:corner}
  \end{center}
\end{figure*}

Our analysis demonstrates that the high-$q$ and low-$q$ likelihood peaks have comparable posterior odds. 
The high-$q$ mode corresponds to a binary black hole system similar to others reported in GWTC-3~\citep{gwtc-3}. 
However, the low-$q$ mode corresponds to a highly spinning, unequal mass ratio event with non-negligible precession (see Fig.~\ref{fig:corner}).

Similar to GW151226, there are other events (e.g., GW190425 and GW190521) with alternative interpretations~\citep{pycbc_ogc_2, pycbc_ogc_3, ias_gw190521}.
\citeauthor{pycbc_ogc_3}'s analysis of GW190425 resulted in a $\xeff$ slightly higher than that reported by \citet{GW190425_LVK}.
However, this is likely due to the differences in the spin priors used in the analyses---when \citeauthor{pycbc_ogc_3} re-weighted their posteriors with a ``low-spin'' prior, they obtained posteriors comparable with those of \citet{GW190425_LVK}. 
In the case of GW190521, while \citet{GW190521}, \citet{ias_gw190521} and \citet{pycbc_ogc_3} report the event to have source frame total mass of $\sim \unit[150]{M_{\odot}}$, they also report different values for the event's mass ratio and spins.
\citet{GW190521} report the event to have an almost equal mass ratio $q\sim0.79^{+0.19}_{-0.29}$ and negligible spin, while \citet{pycbc_ogc_3} and \citet{ias_gw190521} report the event to have a bi-modal {$q$-$\xeff$} distribution, with a low-$q$ peak $(q<0.3, \xeff < 0)$ and a high-$q$ peak $(q>0.3, \xeff \sim 0)$.
These bi-modal posteriors may be due to the different mass and spin priors used in the \citet{GW190521} analysis; by using source-frame mass priors and uniform-in-$\xeff$ prior instead of using detector-frame mass priors and isotropic spin prior. 
Determining if there are other events with multi-modal structures and unaccounted peaks is an area worthy of future study.

An area of future research is to repeat this analysis using a Fisher expansion to estimate the evidence associated with each peak. (This would increase the computational cost by a factor of five as we would need to calculate the likelihood in several points in the vicinity of points $A$ and $B$ to calculate the second derivatives of the likelihood function in the $q$ and $\xeff$ directions, but this would not be prohibitive.) This may yield additional/complementary information beyond what we can tell from comparing the likelihood at just two points.

\section{Acknowledgments}
The authors thank \chia, \mateu\, and \nitz\ for sharing their {$q$-$\xeff$} GW151226 posterior samples. In addition, we thank Shanika Galaudage for their technical assistance.
The authors also thank Horng Sheng Chia and David Keitel for helpful discussions.
Finally, the authors credit the referee for their Fisher expansion idea presented in Section~\ref{sec:discussion} and general improvements to the manuscript.


We gratefully acknowledge the Swinburne Supercomputing OzSTAR Facility for computational resources. All analyses (including test and failed analyses) performed for this study used $\sim100$K core hours on OzSTAR. This would have amounted to a carbon footprint of ${\sim \unit[6.2]{\text{t CO}_2}}$~\citep{greenhouse, energy_to_co2_converter}. However, as OzSTAR is powered by wind energy from Iberdrola Australia, the electricity for computations produces negligible carbon waste.

This material is based upon work supported by NSF's LIGO Laboratory, a major facility wholly funded by the National Science Foundation.
This research uses data from the Gravitational Wave Open Science Center (gw-openscience.org), a service of LIGO Laboratory, the LIGO Scientific Collaboration, the Virgo Collaboration, and KAGRA. LIGO Laboratory and Advanced LIGO are funded by the United States National Science Foundation (NSF) as well as the Science and Technology Facilities Council (STFC) of the United Kingdom, the Max-Planck-Society (MPS), and the State of Niedersachsen/Germany for support of the construction of Advanced LIGO and construction and operation of the GEO600 detector. Additional support for Advanced LIGO was provided by the Australian Research Council. Virgo is funded through the European Gravitational Observatory (EGO), the French Centre National de Recherche Scientifique (CNRS), the Italian Istituto Nazionale di Fisica Nucleare (INFN), and the Dutch Nikhef, with contributions by institutions from Belgium, Germany, Greece, Hungary, Ireland, Japan, Monaco, Poland, Portugal, Spain. The construction and operation of KAGRA are funded by the Ministry of Education, Culture, Sports, Science and Technology (MEXT), Japan Society for the Promotion of Science (JSPS), the National Research Foundation (NRF) and the Ministry of Science and ICT (MSIT) in Korea, Academia Sinica (AS) and the Ministry of Science and Technology (MoST) in Taiwan.

This work is supported by the Australian Research Council (ARC) Centre of Excellence CE170100004.
E. T. acknowledges the support of ARC DP230103088.

\bibliography{refs}

\appendix

\section{Tracking down the source of discrepancies}\label{sec:appendix}
Gravitational-wave astronomers use Bayesian inference to derive posterior distributions of astrophysical parameters given data. These inferences are affected by several assumptions and data analysis choices. Key ingredients include the sampling frequency, the noise power spectral density (PSD) \citep{Littenberg:2015:PhRvD,chatziioannou2019noise,windows,Biscoveanu:2020:PhRvD}, data-cleaning procedures \citep{Pankow2018,Chatziioannou2021}, the prior distribution, and the sampler (and sampler settings) employed to probe the parameter space. All these factors can affect the final posterior distributions. Thus, given two discrepant posterior distributions obtained with the same data, the source of the disagreement may not be immediately evident. However, the deep-followup method presented here can assist us in determining which ingredient(s) are responsible for the difference.

This paper investigates the high-$q$ and low-$q$ posteriors peaks for GW151226. 
We conduct our deep follow-up with the ingredients utilized by the \mateu\ analysis and find slightly more support than \mateu\ for the low-$q$ peak, but not nearly as much as \chia. 
Since we observe a modest difference using the same noise PSD, the same analysis software, etc., it seems plausible---by process of elimination---that the difference is attributable to the sampler's ability (or inability) to probe the low-$q$ peak with the uniform-in-$\xeff$ prior.
As part of ongoing work, we are attempting to test this hypothesis by reanalyzing GW151226 using a uniform-in-$\xeff$ prior, which \cite{ias_gw190521, GW190521_pycbc} suggest probes the parameter space of black hole spins more efficiently than the isotropic spin prior.

However, our findings could have been different; we could have obtained results consistent with \mateu, which would have ruled out the choice of prior as the source of the discrepancy with \chia.
It is instructive to consider what next steps we would take in this hypothetical scenario to understand better the discrepancy between the \chia\ and \mateu\ analyses.
First, to investigate the difference in noise PSD, we would reweight our posteriors samples to see how the results change using the noise PSD from \chia.
The weight for sample $i$ is given by
\begin{align}
    w_i = \frac{{\cal L}(d | \theta_i, \text{PSD}_\text{\chia})}{{\cal L}(d | \theta_i, \text{PSD}_\text{\mateu})} ,
\end{align}
where ${\cal L}$ is the likelihood function of the data $d$ given astrophysical parameters $\theta_i$, and conditioned on one of two different noise models.\footnote{This is an example of importance sampling; see, e.g., \cite{Robert&Casella}. The \mateu\ noise model serves as the proposal distribution, while the \chia\ noise model is what is known as the target distribution.}
If the reweighted samples reproduced key features from \chia, we could attribute the difference to 
\chia's choice of noise PSD.

However, let us imagine that a different choice of noise PSD yields results broadly consistent with \mateu. 
In this scenario, the next step is to repeat the importance sampling procedure but now focusing on a different ingredient, such as the data cleaning procedure.
Once again, we calculate weights, but this time, the target distribution would use the \chia\ data-cleaning procedure while the proposal distribution would use the data-cleaning procedure (if any) from \mateu.
We would continue down the list of data-analysis ingredients until either we find the culprit or run out of plausible explanations for the discrepancy.
If we can find nothing that alters the \mateu\ results to reproduce the salient features from \chia, we may be left to speculate that there is simply a bug in one or more analyses.

\end{document}